\documentclass[12pt]{article}
\usepackage{amsmath}
\usepackage{graphicx}
\usepackage{enumerate}
\usepackage{natbib}
\usepackage{url} % not crucial - just used below for the URL 
\usepackage{color}

\newcommand{\blind}{0}

\usepackage{amssymb,bm,bbm,mathrsfs,subcaption}
\usepackage[english]{babel}
\newtheorem{theorem}{Theorem}
\newtheorem{corollary}{Corollary}

\newcommand{\be}{\begin{equation}}
\newcommand{\ee}{\end{equation}}
\newcommand{\bit}{\begin{itemize}}
\newcommand{\eit}{\end{itemize}}

\def\T{{ \mathrm{\scriptscriptstyle T} }}

\newcommand{\bas}{\begin{eqnarray*}}
\newcommand{\eas}{\end{eqnarray*}}
\newcommand{\ba}{\begin{eqnarray}}
\newcommand{\ea}{\end{eqnarray}}

\newcommand{\bgamma}{\mbox{\boldmath $\gamma$}}

\newcommand{\btheta}{\mbox{\boldmath $\theta$}}

\newcommand{\bSigma}{\mbox{\boldmath $\Sigma$}}

\newcommand{\norm}[1]{\left\lVert#1\right\rVert}

\begin{document}

\def\spacingset#1{\renewcommand{\baselinestretch}%
{#1}\small\normalsize} \spacingset{1}

%%%%%%%%%%%%%%%%%%%%%%%%%%%%%%%%%%%%%%%%%%%%%%%%%%%%%%%%%%%%%%%%%%%%%%%%%%%%%%

\if0\blind
{
  \title{\bf  Locally Sparse Estimation for Simultaneous Functional Quantile Regression}
  \author{Boyi Hu\hspace{.2cm}\\
    Department of Statistics and Actuarial Science, Simon Fraser University\\
    and \\
    Jiguo Cao \\
    Department of Statistics and Actuarial Science, Simon Fraser University}
    \date{}
  \maketitle
} \fi

\if1\blind
{
  \bigskip
  \bigskip
  \bigskip
  \begin{center}
    {\LARGE\bf Locally Sparse Estimation for Simultaneous Functional Quantile Regression}
\end{center}
  \medskip
} \fi

\bigskip
\begin{abstract}
Motivated by investigating how daily temperature affects soybean yield, this article proposes a simultaneous functional quantile regression (FQR) model with a unique twist—a locally sparse bivariate slope function that is indexed by both quantile and time, linked to a functional predictor. The slope function's local sparsity means it holds non-zero values only in certain segments of its domain, remaining zero elsewhere. These zero-slope regions, which vary by quantile, indicate times when the functional predictor has no discernible impact on the response variable. This feature boosts the model's interpretability. Unlike traditional FQR models, which fit one quantile at a time and have several limitations, our proposed method can handle a spectrum of quantiles simultaneously. We tested the new approach through simulation studies, demonstrating its clear advantages over standard techniques. To validate its practical use, we applied the method to soybean yield data, pinpointing the time periods when daily temperature doesn't affect yield. This insight could be crucial for agricultural planning and crop management.
\end{abstract}

\noindent%
\vfill

\newpage
\spacingset{1.75} % DON'T change the spacing!
\section{Introduction}
\label{sec:intro}

Crop yield plays a critical role in agriculture, with profound implications for meeting global demands for food, feed, and fuel. Among the key crops worldwide, soybeans stand out, especially because most of their production—over three-quarters—is used as livestock feed. A much smaller proportion, roughly 7\%, is set aside for conventional soy-based foods like tofu and soy milk. Given this context, understanding how environmental factors affect soybean yield is essential for ensuring stable and sustainable agricultural practices.

Soybeans are a crop that places significant demands on environmental conditions, particularly temperature and water availability \citep{schlenker2009nonlinear,vogel2019effects}. There exists an optimal temperature range for soybean germination and growth, within which small fluctuations don't adversely affect plant development or final yield. This project aims to explore how temperature and water availability impact soybean yield by the end of the growing season. Additionally, we seek to identify the specific periods when temperature is conducive to soybean growth and when it might be too cold or too hot, potentially hindering the plant's development.

To address these problems, we propose a new functional quantile regression model to study the relationship between conditional quantiles of soybean yield and the chosen environmental factors,
\be
Q_Y(u|\bm Z, X) = \bm Z^{\tau}\bm\alpha(u) + \int_0^T\beta(t,u)X(t)dt,
\label{fqr0}
\ee
where $Y$ is a scalar response, $\bm Z = (1, Z_1, \ldots, Z_p)^{\tau}$ is a vector of intercept and scalar predictors, $X(t)$ is a functional predictor defined over $[0, T]$, 
$Q_Y(u|\bm Z, X)$ is the conditional quantile function of $Y$, $u$ is the quantile, and
$\bm\alpha(u) = (\alpha_0(u), \ldots, \alpha_p(u))^{\tau}$ and $\beta(t,u)$ are unknown slope functions.
By focusing on quantile-specific effects, we can pinpoint time periods when temperature has a substantial impact on yield outcomes, beyond traditional mean-based analyses.
For instance, in our soybean yield study, 
$Y$ represents the annual soybean yield, which is the response variable we are interested in modeling and predicting, while
$X(t)$ denotes the daily average temperature over the time interval of interest, which is a functional predictor that varies with time $t$.
We use scalar predictors to capture the water availability of each county. 
Specifically, $Z_1$ is the annual precipitation in each county, and $Z_2$ is the proportion of land that is irrigated for soybean cultivation at the county level.
Regarding the parameters, each entry of the vector $\bm\alpha(u)$ corresponds to a varying coefficient that characterizes the intercept or the influence of a scalar predictor on the $u$-quantile of the response variable $Y$. 
The varying coefficient $\bm\alpha(u)$ allows us to capture how the impact of the scalar predictors changes across different quantiles of the soybean yield distribution.
The function $\beta(t,u)$ is a bivariate slope function indexed by time $t$ and quantile $u$. It describes the dynamic influence of the functional predictor $X(t)$ on the quantiles of the soybean yield $Y$. 

Because there exists a comfortable range of temperature for soybean growth, where small fluctuations of temperature have little influence on soybean yields,
then under model \eqref{fqr0},
$\beta(t,u)$ should be equal to 0 when the temperature at time $t$ is within such comfortable range.
In other words, for a fixed quantile $u$, there exist some time regions $\mathcal M\subset [0, T]$ such that $\beta(t,u) = 0$ for $t \in \mathcal M$, where $\mathcal M$ may vary with the quantile $u$.
This observation serves as a strong motivation for introducing the concept of local sparsity for the functional coefficient $\beta(t,u)$ in the functional quantile regression model \eqref{fqr0}.

The local sparsity refers to the property of having sub-regions within the domain of $\beta(t,u)$ where $\beta(t,u)=0$. 
In other words, there are specific time intervals (sub-regions) during the growing season when the daily average temperature does not affect some quantiles of soybean yield.
These sub-regions correspond to the comfortable temperature range for soybean growth, where temperature fluctuations do not substantially impact the crop's performance for some specific quantiles.
More formally, we assume that there exists a collection of sub-regions $\mathcal{N}\subset [0, T]\times [0,1]$ such that $\beta(t,u)=0$ for all $(t,u)\in \mathcal{N}$.

By incorporating the local sparsity of $\beta(t,u)$ into the model, we can more accurately estimate the dynamic dependence between quantiles of soybean yield and the daily average temperature.
More specifically, identifying the local sparsity region of $\beta(t,u)$ under \eqref{fqr0} allows the model to focus on the periods when temperature has a meaningful influence on soybean yield while disregarding the sub-regions when temperature fluctuations have no effect. It can also provide more precise and meaningful insights about how temperature impacts soybean yield across different quantiles.

Functional quantile regression (FQR) is an important tool for data analysis, particularly in applications related to environmental science. It extends classical quantile regression \citep{koenker1978regression} and enables more flexible and comprehensive modeling of data involving functional predictors. Researchers have developed various extensions of FQR to address different scenarios and accommodate diverse data characteristics.
For instance, \citet{cardot:2005}, \citet{chen:2012}, and \citet{kato:2012} investigated FQR models with a single functional covariate and no non-functional covariates. \citet{lu2014functional} proposed a functional partially linear quantile regression model that incorporates a finite-dimensional scalar covariate. \citet{yu:2016} considered a model with multiple functional covariates and a finite number of scalar covariates. \citet{yao:2017} proposed a partially functional quantile regression model with one functional covariate and high-dimensional scalar covariates. \citet{qingguo2017quantile} studied a semi-parametric functional quantile regression model with both parametric and semi-parametric components of scalar covariates. \citet{ma2019quantile} introduced a functional partially linear model that accommodates multiple functional covariates and ultrahigh-dimensional scalar covariates. More recently, \citet{Beyaztas24} presented a robust estimation procedure for scalar-on-function quantile regression, enabling reliable prediction in the presence of outliers and leverage points. Notably, all of these methods adopt the conventional estimation strategy of fitting separate models for different quantile levels.

Specifically, the conventional estimation strategy of fitting the functional quantile regression model \eqref{fqr0} is to first fix the quantile $u$ and then estimate each entry of $\bm\alpha(u)$ as a scalar parameter, along with estimating $\beta(t,u)$ as a univariate function of $t$ only. 
While this estimation approach provides the benefit of diminishing the dimensionality of the parameter space, it comes with a significant drawback: the lack of smoothness in estimating the bivariate function $\beta(t,u)$.
When the model is fitted for different quantiles separately, the resultant estimator for $\beta(t,u)$ may exhibit smoothness over the time domain concerning $t$, but there is no assurance of smoothness in the direction involving the quantile $u$.
This lack of smoothness can lead to erratic behavior of the estimated $\beta(t,u)$ across quantiles, 
which may not accurately capture the true underlying relationship between the functional predictor and response variable. 

In a recent study, \citet{hu2024simultaneous} proposed a simultaneous estimation strategy for functional quantile regression (FQR) with a single functional covariate and no non-functional covariates. Their approach uses functional principal components—obtained via functional principal component analysis (FPCA) applied to the observed functional covariates—as basis functions to approximate the slope function. The model is then fit jointly across multiple quantile levels, rather than separately. While \citet{hu2024simultaneous} introduced a novel estimation framework for FQR, the model remains relatively simple, and its interpretability is limited.
In practice, the outcome may not be associated with the functional covariate over the entire time domain or across all quantile levels. Therefore, it is crucial to identify the specific regions of time and quantile levels where significant associations exist. However, \citet{hu2024simultaneous} did not provide an inference procedure to detect such regions. Making inference on an unknown bivariate function, such as the slope function $\beta(t,u)$ depending on both time $t$ and quantile index $u$, can be challenging. An alternative strategy is to focus on identifying the regions where no association exists, thereby narrowing down the domain of potential significance. This motivates research on locally sparse estimation for unknown slope functions in functional regression models. Locally sparse estimation refers to strategies that estimate the slope function with sparsity in certain subregions—indicating time-quantile domains where the functional covariate has no effect on the outcome.

Most existing studies exploring local sparsity within the realm of functional data analysis have predominantly focused on functional linear models \citep{Reiss17}, which have demonstrated utility across a range of applications \citep{james2009functional, zhou2013functional}. Incorporating locally sparse estimation into these models can further enhance their interpretability \citep{lin2017locally,fang2020smooth, li2022integrative, gurer2024locally}.
Specifically, \cite{lin2017locally} introduced a functional extension of the smoothly clipped absolute deviation (SCAD) proposed by \cite{fan2001variable}, termed fSCAD. This extension aims to derive locally sparse estimators for univariate scalar-on-function regression. Building on this, \cite{fang2020smooth} and \cite{li2022integrative} subsequently applied fSCAD to address multiple outputs in functional linear regression.
In the context of FQR, \citet{yu2019sparse} and \citet{liang2023locally} investigated locally sparse estimation using different regularization tools to induce sparsity in the slope function. However, these existing methods follow the conventional estimation strategy of fixing the quantile level and estimating the slope function as a univariate function of time. This imposes a significant limitation on the ability to recover the smooth relationship between the conditional distribution of outcome and the functional covariate.

In our proposed method, we use a triangulation based bivariate splines to approximate the nonparametric component $\beta(t,u)$ of interest. Based on that, we develop a novel method to estimate $\beta(t,u)$, which can control the smoothness and guarantee the local sparsity of $\beta(t,u)$. To the best of our knowledge, this article is the first attempt to develop a locally sparse and smooth estimation method for functional quantile regression. Our method revolutionizes the conventional estimation strategies by simultaneously incorporating multiple quantiles into the model fitting process, overcoming the limitation of fixing a single quantile.

To achieve these innovations, our method utilizes roughness regularization to govern the smoothness of the estimation for $\beta(t,u)$ as a bivariate function. Additionally, based on the nature of piecewise bivariate splines defined over a triangulation, we are able to employ penalties on the approximation of $\beta(t,u)$ at the triangle level to attain the locally sparse estimation. The joint application of smoothness control and locally sparse regularization not only enhances the interpretability of the model but also improves the accuracy of estimating $\beta(t,u)$. These enhancements will be demonstrated through the following numerical studies in this article.

The remainder of this article is organized as follows. 
In Section 2, we introduce the model and the corresponding estimator for $\beta(t,u)$. 
In Section 3, we present the main theoretical results.
In Section 4, we first introduce a soybean yield data set and then apply the proposed method to analyze onto it as an illustration.
Section 5 concludes the paper. The computing code for replicating the simulation studies and application in the manuscript is available at \texttt{\url{https://github.com/caojiguo/LocalFQR/}}.

\section{Simultaneous Functional Quantile Regression}
\label{s:model}
For the proposed functional quantile regression model \eqref{fqr0} with locally sparse slope function $\beta(t,u)$, 
we use $\mathscr{U} \subset (0,1)$ to denote the interval containing the quantiles of interest, and use $\Omega = [0, T] \times \mathscr U$ to denote the domain of $\beta(t,u)$ of interest.
Let $S_0 = \{(t,u) \in \Omega: \beta(t,u)=0\}$ denote the null region of $\beta(t,u)$. Let $\hat S_0$ denote the identified $S_0$ using the data.
The next step is to estimate the function $\beta(t,u)$ on the complement set $\Omega\backslash\hat S_0$.
This complement set consists of the regions where the function $\beta(t,u)$ is non-zero, and these regions are of interest.

Estimating the infinite-dimensional functions $\bm\alpha(u)$ and $\beta(t,u)$ directly can be challenging in practice. To address this issue, a common strategy is to use a linear combination of basis functions to approximate these unknown functions and then estimate the coefficients of those basis functions from the data. By assuming $\beta(t,u)$ is smooth enough, we propose to approximate $\beta(t,u)$ using bivariate spline basis functions.
Regarding the bivariate spline basis, there are various choices available for approximation purposes. Two commonly used options are tensor products of B-splines \citep{Stone97} and bivariate Bernstein polynomials over the triangulation \citep{lai:2007,Lai13,Li21}. 
In this article, we choose bivariate Bernstein polynomials over a triangulation for approximating the unknown slope function $\beta(t,u)$ in \eqref{fqr0}.
Each bivariate Bernstein polynomial has its support on a single triangle of the triangulation, and multiple Bernstein polynomials are defined over each triangle.
This local support property ensures that we can use the triangles to estimate the subset $S_0$.

Let $S^r_d(\Delta)$ denote the linear space spanned by bivariate splines defined over a triangulation $\Delta$,
where $r$ is the smoothness condition of this space and $d$ is the degree of the splines.
We use $|\Delta|$ to denote the longest edge of triangles of the triangulation $\Delta$.
Our goal is to find a function $s(t,u) \in S^r_d(\Delta)$ that
can effectively approximate the slope function $\beta(t,u)$ on the domain $\Omega$.
To make our writing and proofs in the subsequent sections more clear,
we use $\{B_j(t,u)\}_{j=1}^{n_B}$ to denote the Bernstein polynomials over the triangulation $\Delta = \{\Lambda_1, \ldots, \Lambda_M\}$,
where $j= 1, \ldots, n_B$ is the index for the polynomials. 
The relationship between $n_B$ and $M$ is $n_B = (d+2)(d+1)M/2$ because $(d+2)(d+1)/2$ Bernstein
polynomials are associated with each triangle of $\Delta$. 
In addition, for each bivariate basis function $B_j(t,u)$,
we denote its support by $\Delta_j$.
In other words, $B_j(t,u)\not=0$ for $(t,u) \in \Delta_j$,
and $B_j(t,u)=0$ for $(t,u) \not\in \Delta_j$. If two Bernstein polynomials $B_j(t,u)$ and $B_k(t,u)$ are associated with
the same triangle, then $\Delta_j$ and $\Delta_k$ are identical.
Figure S1 in the supplementary document is an illustration of six Bernstein polynomials defined over the same triangle when $d=2$.

The function $s(t,u)\in S^r_d(\Delta)$
that approximates $\beta(t,u)$ can be written as a linear combination of Bernstein polynomials 
$\{B_j(t,u)\}_{j=1}^{n_B}$.
Then on the domain $\Omega$,
we have the approximation
\be
\beta(t,u)\approx s(t,u) = \sum_{j=1}^{n_B}\gamma_j B_j(t,u):=\bm B(t,u)^{\tau}\bgamma \in S^r_d(\Delta),
\label{approx4}
\ee
where $\bm B(t,u) = (B_1(t,u), \ldots, B_{n_B}(t,u))^{\tau}$ is the vector of bivariate spline basis functions defined over $\Omega$, and
$\bgamma = (\gamma_1,\ldots,\gamma_{n_B})^{\tau}$ is the corresponding vector of basis coefficients.

To ensure the desired smoothness of the approximation for $\beta(t,u)$, such as continuity and continuity of derivatives, it is necessary to impose linear constraints on the basis coefficients $\bgamma$. These constraints can be expressed as $\bm H \bgamma = 0$, where $\bm H$ is a matrix of linear constraints. By incorporating these constraints, we can enforce the desired smoothness property in the estimation.
However, imposing linear constraints on the coefficients may lead to complications in the optimization procedure.
To overcome this, a common approach is to use QR decomposition to remove the linear constraints,
which is also a commonly used technique of handling identifiability constraints when estimating the generalized additive model \citep{wood2017generalized}.
For a given $\bm H$, by QR decomposition, we have
\be
\bm H^\T=\boldsymbol(\bm Q^*, \bm Q\boldsymbol) 
\begin{pmatrix}
\bm R\\
\bm 0
\end{pmatrix},
\label{smoothconstraint4}
\ee
where $(\bm Q^*, \bm Q)$ is a matrix with orthogonal columns
and $\bm R$ is a upper triangle matrix with nonzero diagonal elements.
Based on the decomposition \eqref{smoothconstraint4}, 
the constraints $\bm{H\gamma=0}$ can be removed by
rewriting $\bgamma$ as $\bgamma =\bm {Q\theta}$,
where $\btheta$ is a vector with lower dimensions than the original parameter vector $\bgamma$.

When the univariate functions in $\bm\alpha(u)$ are assumed to be smooth enough, the entries of $\bm\alpha(u)$ are represented as linear combinations of B-spline basis functions with unknown coefficients: $\alpha_k(u) = \sum_{j=1}^{n_b}\eta_{k,j}b_j(u):=\bm b(u)^{\tau}\bm\eta_k$, where
$\bm b(u) = (b_1(u),\ldots, b_{n_b}(u))^{\tau}$ is the vector of B-spline basis functions (we use $n_b$ to denote the number of B-spline basis functions) defined over $\mathscr{U}$ and $\bm\eta_k = (\eta_{k,1},\ldots,\eta_{k,n_b})^{\tau}$ is the corresponding vector of basis coefficients. Then, the functional quantile regression model (\ref{fqr0}) can be approximated as $Q_Y(u|X,\bm Z)\approx \bm Z^{\tau}\bm V(u)\bm\eta + \int \bm B(t,u)^{\tau}\bm Q\btheta X(t) dt\approx \bm Z^{\tau}\bm V(u)\bm\eta + \bm A(u)^{\tau}\bm Q\btheta$, where $\bm V(u)$ is a block diagonal matrix of B-spline basis function, $\bm\eta = (\bm\eta_0^{\tau},\ldots,\bm\eta_p^{\tau})^{\tau}$ and $\bm A(u)^{\tau} = \int \bm B(t,u)^{\tau} X(t)dt$.
% \begin{align*}
% Q_Y(u|X,\bm Z)&\approx \bm Z^{\tau}\bm V(u)\bm\eta + \int \bm B(t,u)^{\tau}\bm Q\btheta X(t) dt\\
% & = \bm Z^{\tau}\bm V(u)\bm\eta + \bm A(u)^{\tau}\bm Q\btheta,\label{approx4}
% \end{align*}
% where $\bm V(u) = \begin{pmatrix}
%     \bm b(u)^{\tau} & \bm 0 & \cdots \\
%     \bm 0 & \bm b(u)^{\tau} & \cdots\\
%      \cdots& \cdots &\cdots\\
%     \cdots & \bm 0 & \bm b(u)^{\tau}
% \end{pmatrix}$, $\bm\eta = (\bm\eta_0^{\tau},\ldots,\bm\eta_p^{\tau})^{\tau}$ and $\bm A(u)^{\tau} = \int \bm B(t,u)^{\tau} X(t)dt$.

In the proposed estimation procedure, the goal is to estimate the univariate functions $\bm\alpha(u)$ and the bivariate function $\beta(t,u)$ by considering all the quantiles of interest simultaneously. 
Combining multiple quantile regression models in the estimation process has been widely recognized for its advantages such as \cite{zou:2008} and \cite{he2016regularized}.

For a real-valued random variable $Y$, 
the minimizer of $E\{\rho_{u}(Y-u)\}$ is the $u$-quantile of $Y$, 
where $\rho_{u}(x) =x\left(u - \mathbbm{1}\{x < 0\}\right)$ is called the
quantile loss function or check function \citep{koenker1978regression}.
Assume that we observe independent and identically distributed data pairs
$\left\{y_i, \bm z_i,x_i(t)\right\}_{i=1}^n$ as the realizations of $\left\{Y,\bm Z, X(t)\right\}$.
To perform the quantile regression, we consider a set of quantiles of interest denoted by $U \in \mathscr{U}$. 
These quantiles are assumed to be uniformly distributed within the interval $\mathscr{U}$.
The cardinality of $U$, denoted by $n_u$, represents the number of quantiles of interest.
We propose to estimate the basis coefficients by minimizing the following loss function with respect to $\bm\eta$ and $\btheta$:
\be
\frac{1}{nn_u} \sum_{r=1}^{n_u}\sum_{i=1}^{n} 
\rho_{u_r}\left\{y_i - \bm z_i^{\tau}\bm V(u_r)\bm\eta -\bm A_i(u_r)^{\tau}\bm Q\btheta\right\}
\label{sumofquantileloss4}
\ee
where $\bm A_i(u_r) = \int \bm B(t,u_r)^{\tau} x_i(t)dt$.

However, the design matrix involving $\bm z_i$, $\bm b(u_r)$, and $A_i(u_r)$ in (\ref{sumofquantileloss4}) may be ill-conditioned, leading to a highly wiggly estimator for $\beta(t,u)$ when directly minimizing \eqref{sumofquantileloss4}. Consequently, the estimation of $\beta(t,u)$ may not be as smooth as assumed by the model.
To address this issue, a common approach is to add a penalty term to \eqref{sumofquantileloss4} during the minimization procedure. This penalty term can enhance numerical stability and control the smoothness of the estimation for $\beta(t,u)$. One particularly effective tool for this purpose is the roughness penalty.
Detailed discussions on roughness
penalty in the context of functional data analysis can be found in \cite{cardot2003spline}, \cite{Ramsay05}, \cite{Sangalli13}, and \cite{Azzimonti15}.
For a smooth bivariate function $s(t,u)$, the roughness penalty $R(s)$ is defined as
\be
R(s)= \sum_{\Lambda\in\Delta}\int_{\Lambda}
\sum_{d_1+d_2=2}{2 \choose d_1}\left[\nabla^{d_1}_t\nabla^{d_2}_u s(t,u)\right]^2 dtdu.\label{roughness4}
\ee
For any bivariate spline approximation $s(t,u) = \bm B(t,u)^{\tau}\bm Q\btheta$,
the roughness penalty of $s(t,u)$ \eqref{roughness4} can be further written as a quadratic form with a positive semi-definite matrix $\bm G$,
$R(s) = \btheta^{\tau}\bm G\btheta$.
With the help of the roughness penalty $R(\cdot)$,
we propose to estimate the unknown coefficients $\bm\eta$ and $\btheta$ in \eqref{fqr0} by minimizing the following objective function:
\be
\frac{1}{nn_u} \sum_{r=1}^{n_u}\sum_{i=1}^{n} 
\rho_{u_r}\left(y_i - \bm z_i^{\tau}\bm V(u_r)\bm\eta -\bm A_i(u_r)^{\tau}\bm Q\btheta\right)+ \lambda \btheta^{\tau}\bm G\btheta.\label{initial_est}
\ee

\subsection{Locally Sparse Estimation for $\beta(t,u)$}
Our primary objective is to obtain a locally sparse estimator for $\beta(t,u)$, which helps identify the inactive regions of the functional predictor $X(t)$ across different quantiles of the scalar response $Y$. To achieve this goal, we employ a group LASSO method.

By utilizing the triangulation-based spline approximation method, we can impose penalization on the $L_2$-norm of the approximation \eqref{approx4} at the triangle level, which enables us to promote a local sparsity for the target function $\beta(t,u)$ at the level of triangles. 
For this purpose, we propose to incorporate group LASSO penalties into our estimation procedure.
In essence, the group LASSO penalties encourages the bivariate splines over certain triangles to have zero coefficients, which can help identify the inactive regions of the functional predictor $X(t)$ for specific quantiles of $Y$.
{Specifically, we propose the following double-penalized objective function to obtain a smooth and locally sparse estimator of $\beta(t,u)$:
\begin{equation}
\frac{1}{n n_u}\sum_{r=1}^{n_u}\sum_{i=1}^{n} 
\rho_{u_r}\!\left(y_i-\bm z_i^{\tau}\bm V(u_r)\bm\eta-\bm A_i(u_r)^{\tau}\bm Q\btheta\right)
+\lambda_1\,\btheta^{\tau}\bm G\btheta
+\lambda_2\sum_{j=1}^J \|\bgamma_{[j]}\|_2,
\label{grp_LASSO0}
\end{equation}
which combines a roughness penalty and a group LASSO penalty. The resulting bivariate spline-based estimator of $\beta(t,u)$ is smooth while exhibiting sparsity at the triangle (group) level.
As in the standard LASSO, the group LASSO penalty may introduce shrinkage bias in parameter estimation. In practice, we therefore consider an adaptive version that assigns group-specific weights based on an initial estimator, for example the minimizer of~\eqref{initial_est}. The choice of adaptive group LASSO weights is not unique, and several reasonable weighting schemes can be used.}
In this paper, we consider two different types of group LASSO penalty. In the following, we denote the minimizer of \eqref{initial_est} as $(\tilde{\bm\eta},\tilde\btheta)$.

\noindent\textbf{Option 1.} Define
$w_j = \norm{\tilde\bgamma_{[j]}}^{-\eta}_2$,
where $\tilde\bgamma = \bm Q\tilde\btheta$ and $\tilde\bgamma_{[j]}$ denote the entries of $\tilde\bgamma$ corresponding to the triangle $\Delta_j$.
We can minimize the following objective function to obtain estimates of $\bm\eta$ and $\btheta$,
\be
\frac{1}{nn_u} \sum_{r=1}^{n_u}\sum_{i=1}^{n} 
\rho_{u_r}(y_i - \bm z_i^{\tau}\bm V(u_r) \bm\eta -\bm A_i(u_r)^{\tau}\bm Q\btheta)+ \lambda_1 \btheta^{\tau}\bm G\btheta
+ \lambda_2\sum_{j=1}^Jw_j\norm{\bgamma_{[j]}}_2,
\label{grp_LASSO1}
\ee
where $\bgamma_{[j]}$ denotes the entries of $\bgamma$ corresponding to the triangle $\Delta_j$.

\noindent\textbf{Option 2.} 
Denote the initial estimate for $\beta(t,u)$ as $\tilde\beta(t,u)= \bm B(t,u)^{\tau}\tilde\bgamma$.
Define $w_j = \norm{\tilde\beta|_{\Delta_j}}^{-\eta}_2$,
where $\tilde\beta|_{\Delta_j}$ is the restriction of $\tilde\beta(t,u)$ to the triangle $\Delta_j$.
Another way to obtain reasonable estimates on $(\bm\eta,\btheta)$ is to minimize the following objective function,
\be
\frac{1}{nn_u} \sum_{r=1}^{n_u}\sum_{i=1}^{n} 
\rho_{u_r}(y_i - \bm z_i^{\tau}\bm V(u_r) \bm\eta -\bm A_i(u_r)^{\tau}\bm Q\btheta)+ \lambda_1 \btheta^{\tau}\bm G\btheta
+ \lambda_2\sum_{j=1}^Jw_j\norm{f_{\btheta}|_{\Delta_j}}_2,
\label{grp_LASSO2}
\ee
where $f_{\btheta}(t,u) = \bm B(t,u)^{\tau}\bm Q\btheta$, and $f_{\btheta}|_{\Delta_j}$ is the restriction of $f_{\btheta}$ to $\Delta_j$.

For both versions of the group-LASSO penalties, the corresponding objective function to be minimized consists of the sum of quantile losses across multiple quantile levels and two regularization terms. The first regularization term is a roughness penalty, expressed in quadratic form, which controls the smoothness of the estimated bivariate slope function. The second is a group-LASSO penalty that promotes local sparsity in the estimated slope surface. This minimization problem falls into the category of second-order cone programming (SOCP) and can be formulated as a convex optimization problem that adheres to the Disciplined Convex Programming (DCP) ruleset. As a result, the problem is solvable using the R package \texttt{CVXR} \citep{fu2020cvxr}, with either the \texttt{SCS} or \texttt{ECOS} solver.

{An important practical issue in quantile regression is the avoidance of
crossing conditional quantile estimates, since a valid conditional quantile
function must be non-decreasing in the quantile index $u$
\citep{hu2024simultaneous, neocleous2008monotonicity}.
However, when quantiles are estimated separately or under limited sample sizes,
there is generally no guarantee that the estimated conditional quantile
$\widehat Q_Y(u \mid X_i, \bm Z_i)$ is non-decreasing in $u$ for all subjects
$i=1,\ldots,n$.
Several approaches have been proposed in the literature to address this issue.
In \citet{schnabel2013simultaneous}, the authors considered a setting in which the
conditional quantile curve of a scalar response is indexed by a single scalar
covariate. By treating the quantile index $u$ as an additional covariate, they
proposed estimating the conditional quantile function as a smooth surface over
the covariate and $u$, with asymmetric difference penalties applied to encourage
monotonicity in the quantile direction.
\citet{takeuchi2006nonparametric} proposed estimating multiple quantiles
simultaneously while imposing explicit non-crossing constraints at the observed
data points.
In \citet{liu2011simultaneous}, the authors adopted a kernel-based representation
for the quantile regression function and enforced non-crossing by imposing linear
inequality constraints on the kernel coefficients.
In this paper, we adopt the general idea of \citet{takeuchi2006nonparametric} and
incorporate monotonicity constraints directly into the optimization procedures
defined in \eqref{grp_LASSO1} and \eqref{grp_LASSO2}.} Specifically, we can first select a set of quantiles of interest, denoted by $S$. Then, we impose inequality constraints to ensure monotonicity across quantiles: for any $\tau_1 > \tau_2$ with $\tau_1, \tau_2 \in S$, the estimated $\tau_1$-quantile should be no smaller than the estimated $\tau_2$-quantile for each individual $i$. This approach helps prevent conditional quantile estimates from crossing, thereby preserving the monotonicity property of the true quantile functions.
However, this constrained optimization becomes computationally expensive when the sample size is large, as the number of constraints grows with both the sample size and the number of quantile levels considered. Moreover, solving the resulting problem requires the use of the commercial solver \texttt{MOSEK} \citep{aps2019mosek} under the disciplined convex programming framework. Developing a more efficient and user-friendly algorithm for handling such constraints remains an open challenge. For these reasons, we did not incorporate the non-crossing quantile constraint into the current framework and leave it as a direction for future work.

\subsection{Parameter Tuning}
In this subsection, we present how to determine
the values of the tuning parameter $\lambda$ \eqref{initial_est}, $\lambda_1$ and $\lambda_2$ in \eqref{grp_LASSO1} or \eqref{grp_LASSO2}.
The weights $w_j$ are essential for the group LASSO penalties used to achieve local sparsity in the estimation of $\beta(t,u)$, and are usually determined based on some initial estimates.
In our framework, we propose to use the estimation method corresponding to the objective function \eqref{initial_est} to obtain an initial estimate for $\btheta$. 
For this initial estimate, it only depends on the tuning parameter $\lambda$, and it does not possess any locally sparsity feature as there is no sparsity regularization involved in the function \eqref{initial_est}.
For this step, we choose to use a 10-fold cross validation to decide the optimal value for $\lambda$, and then obtain the initial estimate for $\btheta$.
Based on the initial estimate of $\btheta$, we compute the weights $w_j$ either based on \eqref{grp_LASSO1} or \eqref{grp_LASSO2}.

The next step is to determine the values for $\lambda_1$ and $\lambda_2$ in the objective functions \eqref{grp_LASSO1} or \eqref{grp_LASSO2}. To achieve this, for a given set of tuning parameters, we first use the entire dataset to fit the model based on the objective function in \eqref{grp_LASSO1} or \eqref{grp_LASSO2}, thereby identifying an active set of triangles from the triangulation. Specifically, we identify the triangles with non-zero coefficients based on $\hat\bgamma$, obtained by minimizing \eqref{grp_LASSO1} or \eqref{grp_LASSO2}, where $\hat\bgamma = \bm{Q} \hat{\btheta}$.
Then, in each round of cross-validation, we only “refit” parameters within the active set of triangles, excluding the group LASSO penalties, to compute the validation loss. This approach helps to avoid potential bias introduced by the group LASSO penalty in the estimation of $\bgamma$.

To make sure that the identified inactive regions of the functional predictor (triangles with zero coefficients) are properly excluded from the model fitting, we impose additional linear constraints on $\bgamma$ during the estimation process, ensuring that the coefficients of the selected triangles are set to zero. These additional linear constraints on $\bgamma$ can be expressed as: $\bm D\bgamma = \bm 0$,
where the matrix $\bm D$ is filled with 0's and 1's.
This ensures that the basis coefficients corresponding to the selected triangles with zero coefficients remain zero in the subsequent model fitting steps.

{In practice, we prefer to use smaller candidate values for the roughness penalty. Although smaller values may increase the variability of the slope estimates and potentially result in slightly wiggly functions, they help reduce estimation bias. Therefore, we do not recommend using very large values. This consideration is particularly important when both a roughness penalty and a group LASSO penalty are incorporated into the estimation procedure. In such cases, a large roughness penalty can lead to overly sparse estimates, especially near the boundaries between signal and non-signal regions. Because the curvature on these boundaries tends to be higher, a large roughness penalty can cause over-penalization in these regions, which may suppress true signals and distort the estimated function.}

\section{Theoretical Properties}
Define 
\[
\Gamma_Xf = \int_s \text{cov}(s,t)f(s,u)ds, \quad \|f(t,u)\|_{\Gamma_X,2} = \left(\int_u\int_t\Gamma_Xf\times fdt du\right)^{1/2}.
\]
To derive the asymptotic property of the proposed estimator $\hat\beta(t,u)$ by minimizing \eqref{initial_est}, we make the following assumption:
\begin{enumerate}
  \item[(A1)] $\left\{Y_i, \bm Z_i, X_i(t)\right\}_{i=1}^n$ are i.i.d.
  \item[(A2)] $\norm{X(t)}\leq C_0 < \infty$ a.s.
  \item[(A3)] The functions in $\bm\alpha(u)$ are supposed to have $p'$th derivatives $\bm\alpha^{(p')}(u)$ such that
  \[
  \left|\alpha_k^{(p')}(t)-\alpha_k^{(p')}(s)\right|\leq C\left|t-s\right|^{\vartheta}, \quad s,t\in (0,1),
  \]
  where $C_1 >0$ and $\vartheta\in[0,1]$.
  In what follows, we set $\nu = p' + \vartheta$.
  \item[(A4)] The eigenvalues of $\Gamma_X$ are strictly positive.
  \item[(A5)] $\beta(t,u) \in W^{d+1}_q(\mathscr T \times \mathscr U)$.
  \item[(A6)] The random variable $\epsilon_i$ defined by $\epsilon_i = Y_i - \bm Z_i^{\tau}\bm\alpha(u) - \int\beta(t,u)X_i(t)dt$ has a density function $f$ continuous and bounded below by a strictly positive constant at $0$.
\end{enumerate}
\setcounter{theorem}{0}
\begin{theorem}
Under the assumptions A1-A6, suppose $n_b \sim |\Delta|^{-(d+1)/\nu}$, and $n_u$ is large enough, then
\[
\|\hat\beta(t,u) - \beta(t,u)\|_{\Gamma_X,2} = O_p\left(|\Delta|^{d+1} + \dfrac{1}{n\lambda|\Delta|^2} +\lambda\right),
\]
except on a set whose probability goes to zero as $n$ goes to infinity.
\end{theorem}

{
We next establish a non-asymptotic upper bound for the $l_2$ estimation error of
$\hat\beta(t,u)$ obtained from \eqref{grp_LASSO0}. Without loss of generality, we assume that the covariate component $\bm Z$
in \eqref{fqr0} consists only of an intercept term. 
For notational simplicity, now we use $\bgamma$ to denote both parameter vectors associated with B-splines and Bernstein polynomials. That is, the first $n_b$ entries of $\bgamma$ are associated with B-splines and the remaining $n_B$ entries are associated with Bernstein polynomials over the triangulation.
With this notation, we let $\bgamma_0$ denote the first $n_b$ entries and $\bgamma_{-1}$ denote the remaining $n_B$ entries.
Define $\bm {\tilde w}_{i,u} = \left[\bm z_i^\top\bm V(u), \bm A_i(u)^\top\right]$,
$\bm{\tilde W}_u = [\bm {\tilde w}_{i,u}, \ldots, \bm {\tilde w}_{i,u}]^\top$, and $\hat\bSigma:=n^{-1}\bm{\tilde W}_u^\top\bm{\tilde W}_u$.
Define $\kappa\in (0,1)$ by
\[
1-\kappa = \text{P}\left(\|\hat\bSigma_j-\bm I_{|G_1|}|\|\leq 0.5, \quad\forall 0\leq j\leq J, \quad\forall u\in U\right).
\]
The numerical constant $0.5$ is not essential and can be replaced by any fixed $c\in(0,1)$. 
This condition defines a high-probability event whose complement has probability $\kappa$. 
When $\kappa$ is small, it implies that, uniformly over $j$ and $u$, the sample covariance matrices 
$\hat{\bSigma}_j$ are well conditioned and remain close to their population counterparts.
We further assume:
\begin{enumerate}
\item [(A7)]  
(Restricted eigenvalue condition). $\phi_{\min}:=\phi_{\min}(c_0, \bar S):=\inf_{\bm a\in\mathbbm C\cap \mathbbm S^{n_b+n_B-1}}\min_{u\in U}\norm{\bm\Sigma_u^{1/2}\bm a}>0$, where $\bm\Sigma_u$ is the covariance matrix of design matrix after spline approximations at quantile $u$ and $U$ is the chosen candidate set of quantiles. We also define $\phi_{\max}:=\phi_{\max}(c_0, \bar S):=\sup_{\bm a\in\mathbbm C\cap \mathbbm S^{n_b+n_B-1}}\max_{u\in U}\norm{\bm\Sigma_u^{1/2}\bm a}$. $\phi_{\min}$ and $\phi_{\max}$ both depend on the construction of triangulation.
\end{enumerate}

Take a sparse vector $\bar\bgamma_{-1} \in \mathbbm R^{n_B}$ such that $\bar\bgamma_{-1}$ satisfies the linear constraints $\bm H\bar\bgamma_{-1} = \bm 0$ and the corresponding approximation error for $\beta(t,u)$ is bounded by some small constant $a_{\Delta}$:
\[
\sup_{(t,u)\in\Omega}|\bar\beta(t,u) -\beta(t,u)|\leq a_{\Delta}.
\]
Furthermore, take another vector $\bar\bgamma_0$ such that the B-splines approximation error for the intercept function $\alpha(u)$ is also bounded by $a_{\Delta}$.
Define $\bar\bgamma$ as: $\bar\bgamma = (\bar\bgamma_0^\top, \bar\bgamma_{-1}^\top)^\top$.
Let \(C_{\bar\beta}\) be the value of the roughness penalty of
\[
\bar\beta(t,u) := \bm B(t,u)^\top \bar\bgamma_{-1}.
\]
For some $h>0$, define
\begin{align*}
r_* &:= 4C_fa_{\Delta}/c_f\phi_{\min}\vee hC_{\bar\beta},\\
\epsilon_{A_3}^* &:=
\sqrt{3}\,(2\rho_{\max}(\bm G)C_{\bar\bgamma}\lambda_1+\lambda_2)\sqrt{J}
+\frac{\sqrt{6}\,p_{\Delta}^{1/2}c_4^{1/2}\sqrt{J}}{\sqrt{n}}
+\frac{\sqrt{6}\,p_{\Delta}^{1/2}c_4 A_3\sqrt{J\log J}}{\sqrt{n}},
\end{align*}
where $\rho_{\max}(\bm G)$ indicates the largest eigenvalue of $\bm G$.
\begin{theorem}
Assume conditions (A1)-(A7).
Let $A_1$, $A_2$ and $A_3$ be any positive numbers.
We choose tuning parameters $\lambda_1$ and $\lambda_2$ such that $n_U\left((c_0-1)\lambda_2-h^{-1}\lambda_1\right) > 2\sqrt{2}n^{-1/2} +  1.5C_f a_{\Delta} + A_1n^{-1/2}+A_2\sqrt{\log J/n}$.
With probability at least $1-2e^{-A_1^2} - 16Je^{-A_2^2/128}-64J^{1-A_3^2}-\kappa$, for $\hat\gamma$ obtained by minimizing \eqref{grp_LASSO0}, we have
\[
\|\hat\bgamma - \bar\bgamma\|_2 \leq \frac{4}{c_f \phi_{\min}^2}
\left[\epsilon_{A_3}^*\vee
\left(
\frac{24(c_0+1)\sqrt{2 p_{\Delta} J}}{\sqrt{n}}
\;\vee\;
\frac{2\sqrt{2}\,\phi_{\max}}{\sqrt{n}}
\right)
\right]\vee r_* \,.
\]
\end{theorem}
A detailed proof for the theorems in this section are available in the supplementary material.

To gain further intuition on the error bound, we consider the corresponding
asymptotic situation.
In Theorem 2, suppose the tuning parameters $\lambda_1$ and $\lambda_2$ are chosen such that $2\rho_{\max}(\bm G)C_{\bar\bgamma}\lambda_1\lesssim n_u^{-1}n^{-1/2}\sqrt{\log J}$, and $\lambda_2\lesssim n_u^{-1}n^{-1/2}\sqrt{\log J}$. Note that $\rho_{\max}(\bm G) \sim |\Delta|^{-2}$ and $C_{\bar\bgamma}\sim |\Delta|^{-1}$. Under this choice, the leading term in $\epsilon_{A_3}^*$ is $\frac{\sqrt{6}\,p_{\Delta}^{1/2}c_4 A_3\sqrt{J\log J}}{\sqrt{n}}\sim \sqrt{J\log J}n^{-1/2}$. Regarding the term $r_*$, the upper bound of approximation errors $a_{\Delta}$ is of order $O(|\Delta|^{d+1})$ and the number of triangles $J\sim |\Delta|^{-2}$. If we choose $h\sim |\Delta|^{d+1}$, then $r_*\sim |\Delta|^{d+1}$. In addition, we impose the following assumptions:
\begin{enumerate}
\item[(A8)] $1\lesssim \phi_{\min}\leq\phi_{\max}\lesssim 1$.
\item[(A9)] $\kappa\to\infty$ as $n\to\infty$.
\end{enumerate}
Then we can derive the following corollary.
\begin{corollary}
Assume conditions (A1)-(A9). Take $\lambda_1\sim n_u^{-1}n^{-1/2}\sqrt{\log J}|\Delta|^3$, and $\lambda_2\sim n_u^{-1}n^{-1/2}\sqrt{\log J}$ such that $n_U\left((c_0-1)\lambda_2-h^{-1}\lambda_1\right) > 2\sqrt{2}n^{-1/2} +  1.5C_f a_{\Delta} + A_1n^{-1/2}+A_2\sqrt{\log J/n}$ for some $h\sim |\Delta|^3$. 
Then as $n\to\infty$, we have
\[
\|\hat\bgamma - \bar\bgamma\|_2 = O_p\left(|\Delta|^{-1}(-2\log|\Delta|)^{-1/2}n^{-1/2}+|\Delta|^3\right).
\]
\end{corollary}
By \cite{lai:2007}, under the conditions of Corollary 1, for some constant $C$ we have
\[
|\hat\beta(t,u)-\bar\beta(t,u)|_2\leq C|\Delta|\|\hat\bgamma - \bar\bgamma\|_2 = O_p((-2\log|\Delta|)^{-1/2}n^{-1/2}+|\Delta|^4).
\]
}

\section{Real Data Analysis}
Temperature plays a crucial role in soybean growth during all stages, including the vegetative and reproductive phases. 
Both high and low temperatures can have adverse effects on soybean germination and growth. Therefore, understanding the relationship between daily temperature and soybean yield at different quantiles is vital for gaining a comprehensive understanding of soybean growth dynamics.

In this study, we aim to investigate the influence of daily temperatures on soybean yield across various quantiles while controlling for other environmental factors. 
By accounting for factors such as annual precipitation and the ratio of irrigated area, we can better understand the independent contribution of temperature to soybean yield. Quantile regression allows us to analyze the relationship between variables at different points of the distribution, going beyond traditional mean-based regression analysis. By examining multiple quantiles, we can identify how temperature affects soybean yield under various conditions, including both favorable and unfavorable scenarios.

The data collected from the United States Department of Agriculture (USDA) website and the National Oceanic and Atmospheric Administration (NOAA) website provides valuable information on soybean yield and related environmental variables in Kansas between 1991 and 2006. The data covers an important agricultural period and geographical region, as Kansas is one of the leading states in soybean production. 
% \begin{figure}[htbp]
% \centering
%     \includegraphics[width=12cm]{raw_avg_temp.pdf}
% \caption{A sample of daily average temperature curves of counties in Kansas. The unit of the y-axis is the Celsius temperature scale.}
% \label{raw_temp_data4}
% \end{figure}

\begin{figure}[htbp]
    \centering
    \includegraphics[width = 10cm]{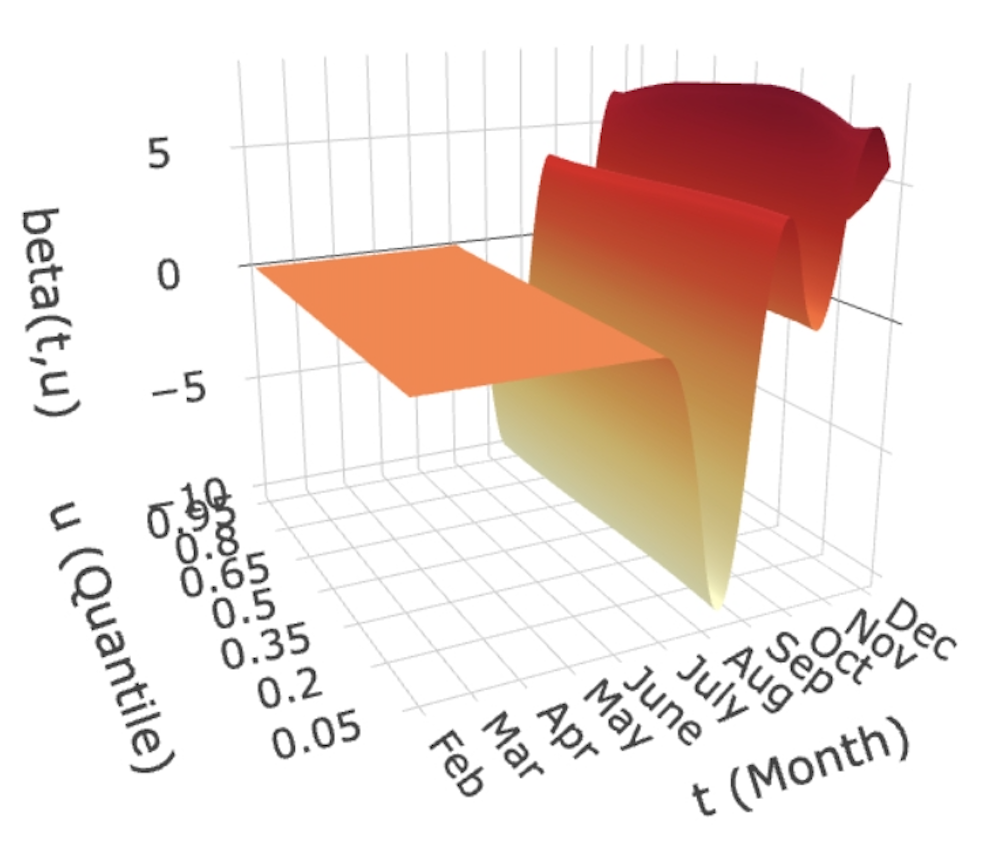}
    \caption{The Estimated slope function $\beta(t,u)$ for the simultaneous functional quantile regression model \eqref{soybean_model} based on the Kansas soybean yield data set.}
    \label{soybean_beta}
\end{figure}

In the following analysis, we consider three predictors for investigating the annual soybean yield of each county in Kansas. 
The functional predictor, denoted as $X(t)$, represents the daily average temperatures. We specifically consider the temperature data from February to November, allowing us to investigate the influence of temperature during both the cold and warm months on soybean yield. 
% Figure \ref{raw_temp_data4} displays a sample of the daily average temperature measurements of Kansas.
The two scalar predictors, annual precipitation and the ratio of irrigated area to harvested land (Z2), denoted as $Z_1$ and $Z_2$, provide additional environmental information that can impact soybean growth.
Finally, the response variable, denoted as $Y$, represents the annual soybean yield per bushel for each county in Kansas.
Then the model we want to investigate is as follows:
\be
Q_Y(u|X(t),Z_1,Z_2) = \alpha_0(u) + 
\alpha_1(u)Z_1+ \alpha_2(u)Z_2 + \int_{\mathcal{T}}\beta(t,u)X(t)dt.\label{soybean_model}
\ee
We fit the model with 19 quantiles uniformly distributed between 5\% and 95\%. Figure \ref{soybean_beta} displays the estimated slope function $\hat\beta(t,u)$ using the proposed simultaneous estimation method.

% \begin{figure}[htbp]
%     \centering
%     \includegraphics[width = 12cm]{soybean_beta_cardot.png}
%     \caption{The Estimated slope function $\beta(t,u)$ using the method proposed in \cite{cardot:2005} based on the Kansas soybean yield data set.}
%     \label{soybean_beta_cardot}
% \end{figure}

% Figure \ref{soybean_beta} displays the estimated slope function $\hat\beta(t,u)$ using the proposed simultaneous estimation method, and Figure \ref{soybean_beta_cardot} presents the estimated $\hat\beta(t,u)$ using the conventional single quantile based estimation method developed by \cite{cardot:2005}.
% Comparing these two figures, we can observe that the estimate obtained from proposed method is much smoother than that estimated by  the method of \cite{cardot:2005}, which is one of the major advantages of our method compared to the conventional method.
% In addition, several trends are evident in Figure \ref{soybean_beta}, which cannot be directly and clearly observed in Figure \ref{soybean_beta_cardot}.

{In Figure \ref{soybean_beta}, the estimated effect function $\hat\beta(t,u)$ reveals distinct time-varying associations between daily average temperature in Kansas and soybean yield across quantiles. Prior to June, temperature appears to have no substantial impact on yield across all quantiles, suggesting that early-season temperatures are generally suitable for soybean germination and initial growth. From July through September, $\hat\beta(t,u)$ is consistently negative for the 5\% to 95\% quantiles, indicating that lower temperatures during the late summer are beneficial to soybean development—likely due to reduced heat stress during critical growth stages such as pod filling.
From late September to mid-October, $\hat\beta(t,u)$ transitions to positive values across all quantiles, implying that higher temperatures during this early fall period may promote continued growth and maturation. However, beginning in mid-October, the slope function declines and becomes negative toward the end of the month, suggesting that unseasonably warm conditions during this time may hinder physiological maturation or dry-down processes. Interestingly, starting in November, $\hat\beta(t,u)$ rises again and remains positive, indicating that warmer temperatures in November are associated with higher yields.
A plausible explanation is that in years with warmer Novembers, soybean crops—particularly late-planted varieties or second-crop systems (e.g., soybeans following wheat)—are less likely to experience early frost damage, allowing for more complete maturation and improved seed quality. These findings highlight the importance of temperature timing on soybean yield and underscore the sensitivity of late-season development to climate variability.}

% In addition, Figure \ref{quantileNov15} provides another illustration of $\hat\beta(t,u)$ varying with quantiles when the time $t$ is chosen as November 15th.
% We can observe that $\hat\beta(t,u)$ is positive for all the quantiles between 5\% and 95\%, emphasizing that a higher temperature in this time period is beneficial of soybean growth and yield.
% Moreover, the influence of the winter temperature on the lower quantiles of the soybean growth and yield seems to be stronger than that on the upper quantiles of the soybean growth and yield.
% Additionally, Figure \ref{quantileNov15} also further demonstrates the smoothness of the estimation for $\beta(t,u)$ in the direction of quantiles based on the proposed method.
% \begin{figure}[htbp]
%     \centering
%     \includegraphics[width = 10cm]{quantileNov15.eps}
%     \caption{The estimated slope function $\beta(t,u)$ varying with quantiles when the time $t$ is fixed as November 15th for the simultaneous functional quantile regression model \eqref{soybean_model} using the Kansas soybean yield data set.}
%     \label{quantileNov15}
% \end{figure}

Overall, the estimate $\hat\beta(t,u)$ provides valuable insights into the dynamic relationship between daily average temperature and soybean yield in Kansas.
Specifically, 
our findings highlights that the temperature in Kansas before June is optimal and comfortable for soybean germination and growth.
The late summer from July to September is one critical time period,
and a lower temperature during this period is favorable for the soybean growth. 
The fall and early winter is another important time period, and during this period, a higher temperature is advantageous for the soybean growth.
These finds can aid in developing strategies to optimize soybean production in the target region.

\section{Conclusions and Discussion}
\label{sec:conc}

In this article, we introduce a novel approach for estimating the bivariate slope function $\beta(t,u)$ under the functional quantile regression model \eqref{fqr0}.
The goal is to investigate and understand the dynamic relationship between functional predictors and scalar response variables.
Our proposed method employs a locally sparse estimation technique, which allows us to identify specific domains of interest when the influence of the functional predictor on the response variable is significant.

We use simulations to compare the performance of the estimation for the slope function under FQR model based on the proposed method and the conventional method.
The numerical results provide evidence to demonstrate the superiority of the proposed method.
Additionally, we further use a soybean yield data set to illustrate the difference between the proposed method and the conventional. The real data results demonstrate that the proposed method can improve the model interpretability.

{Compared to another commonly used locally sparse estimation emthod, fSCAD, the main advantage of the proposed method lies in its computational simplicity. Unlike group lasso, the fSCAD regularization is non-convex, and its numerical optimization typically relies on local linear or local quadratic approximations to iteratively approximate the non-convex objective with convex surrogates. These iterative strategies require careful specification of the initial values and step sizes. Poor choices of initial points can lead to non-convergence, while inappropriate step sizes can cause inefficiency or suboptimal solutions: small step sizes may result in slow convergence, whereas large step sizes may cause the algorithm to converge to a local, rather than the global optimum. In contrast, group LASSO penalties do not suffer from these issues, as they lead to convex optimization problems that are more stable and computationally efficient to solve.}

\bigskip
\begin{center}
{\large\bf SUPPLEMENTARY MATERIAL}
\end{center}

\begin{description}

\item[FQR\_Supplementary.pdf:] the supplementary document containing simulation studies and the proofs for the theoretical results in the manuscript. 

\item[Data availability statement:] The data that support the findings of this study are openly available at \texttt{https://anonymous.4open.science/r/LocalFQR-1322/}.

% \item[FQR\_CODE.zip:] the computing code for replicating the simulation studies and application. 

\end{description}

% \begin{thebibliography}{chicago}
\bibliographystyle{apalike}   
\bibliography{Bibliography-SFQR,References,bibfile} 
\end{document}